\documentclass[12pt]{article}
\pdfoutput=1
\usepackage{geometry,enumerate,amsmath,amssymb}
\usepackage{fullpage}
\usepackage{graphicx}
\usepackage{bm}
\numberwithin{equation}{section}

\newcommand{\be}{\begin{equation}}
\newcommand{\ee}{\end{equation}}
\newcommand{\bea}{\begin{eqnarray}}
\newcommand{\eea}{\end{eqnarray}}

\renewcommand{\epsilon}{\varepsilon}

\begin{document}
\title{
  Skyrmions in models with pions and rho mesons
}
\author{
  Carlos Naya and Paul Sutcliffe\\[10pt]
 {\em \normalsize Department of Mathematical Sciences,}\\
 {\em \normalsize Durham University, Durham DH1 3LE, United Kingdom.}\\ 
{\normalsize Email: carlos.naya-rodriguez@durham.ac.uk, \ p.m.sutcliffe@durham.ac.uk}
}
\date{April 2018}

\maketitle
\begin{abstract}
  A problem with the standard Skyrme model is that Skyrmion binding energies are around $15\%$, being much larger than the order $1\%$ binding energies of the nuclei that they aim to describe. Here we consider theories that extend the standard Skyrme model of pions by including rho mesons, via dimensional deconstruction of Yang-Mills theory with an extra dimension.
  We report the first results of parallel numerical computations of multi-Skyrmions in   theories of this type, including a model that reduces Skyrmion energies below those of the standard Skyrme model whilst retaining exactly the same Faddeev-Bogomolny energy bound.
  We compute all Skyrmions with baryons numbers up to 12 and find that the inclusion of rho mesons reduces binding energies to less than $4\%,$ and therefore moves Skyrmion theory closer to experimental data. Furthermore, we find that this dramatic reduction in binding energies is obtained without changing the qualitative features of the Skyrmions, such as their symmetries.
  \end{abstract}

\newpage
\section{Introduction}\quad
Skyrmions are topological solitons that model nuclei, with the number of solitons being identified with the baryon number of the nucleus (for reviews see \cite{MS,BR}).
The standard Skyrme model \cite{Sk} reproduces some physical properties of nuclei with reasonable success, but a major problem is that it yields classical Skyrmion binding energies \cite{BS-full} of around $15\%$, and therefore an order of magnitude greater than typical nuclear binding energies that are more like $1\%$ of the nuclear mass.

The only fields that are included in this simplest version of the theory are pions, these being the lightest mesons, and indeed they are often assumed to be massless.
A natural approach to reducing the tight binding of Skyrmions is to incorporate the heavier mesons that are neglected within the effective nonlinear pion theory. In fact there is a long history of attempts to include the next lightest meson, the rho meson, within the Skyrme model \cite{Ad,MZ,BKUYY,BKY,HY} but this introduces a host of extra difficulties. In particular, the inclusion of rho mesons significantly increases both the number of fields in the model and the number of terms in the Lagrangian, making numerical computations of Skyrmions considerably more demanding and at the limit of current capabilities. Furthermore, there are a large number of unknown coupling constants that make it impractical to simply search within this space of theories for a parameter set that might reduce Skyrmion binding energies.

A potential solution to the problem of fixing the coupling constants has been proposed \cite{Sut} by using dimensional deconstruction of pure Yang-Mills theory in one higher dimension to yield a Skyrme model of pions coupled to any number of vector mesons: this number being set by the truncation level in a basis expansion. This approach has the considerable advantage that all coupling constants are automatically determined and is expected to reduce binding energies, because in the limit in which the truncation level tends to infinity the theory flows to a BPS limit where binding energies vanish. Furthermore, the coupling constants are related in such a way that a topological lower bound on the energy survives the inclusion of any number of vector mesons. The level one truncation provides a Skyrme model of pions coupled to rho mesons but this theory is sufficiently
complex that to date only the single Skyrmion has been computed in this model, so multi-Skyrmion solutions and their binding energies are unknown.

By applying an extension of the Atiyah-Manton \cite{AM} instanton holonomy approximation of the pion Skyrme field to include rho mesons too, approximate Skyrmions have been calculated for baryon numbers up to four \cite{Sut2}, and the results support the conjecture that binding energies are substantially reduced. However, in addition to approximating both the pion and rho meson fields, this approach also makes the assumption that the symmetries of Skyrmions in the extended theory are identical to those in the standard Skyrme model. Moreover, this method is difficult to apply beyond baryon number four because the assumed symmetries of the Skyrmion are not enough to uniquely determine the Yang-Mills instanton that is required in this approximation. Nonetheless, these approximate results are sufficiently encouraging to warrant a full numerical solution of multi-Skyrmions within this model of pions and rho mesons. Here, we present the results of parallel numerical computations of Skyrmions for all baryons numbers from one to twelve and find that binding energies are dramatically reduced from the $15\%$ in the standard Skyrme model to less than $4\%$. Furthermore, we find that this significant reduction in binding energies is achieved without inducing any changes to the qualitative features of the Skyrmions, such as their symmetries, that are responsible for some of the current successes of the Skyrme model.

We also consider some properties of the most general theory of pions and rho mesons that can be obtained by our dimensional deconstruction approach upon varying the basis functions. This analysis reveals that the rho mesons in these models can never be written in the form of a massive Yang-Mills field, but does highlight one particular theory where the same Faddeev-Bogomolny energy bound of the standard Skyrme model survives the inclusion of rho mesons. We compute Skyrmions in this model for all baryons numbers from one to twelve and find similar results to the previous theory, with binding energies once again reduced to $4\%$, indicating some rigidity of results in models of this type. 

Before discussing our new results, it is worth pointing out that there have been some recent alternative approaches to modifying the standard Skyrme model to address the issue that Skyrmions are too tightly bound. There are two different theories that are both based on the idea of a suppression of the standard leading order term in the Lagrangian in a derivative expansion of the pion field. The first of these is a BPS Skyrme model \cite{Adam2010a,Adam2010b}, including only a term of sixth order in derivatives and a potential term. As binding energies vanish in this BPS model then it provides a good starting point to approach small binding energies \cite{Adam2013a,Adam2013b}. A problem with the BPS model is 
an embarrassment of riches, in that there is an infinite symmetry group that allows Skyrmions to take arbitrary shapes. It is hoped that perturbing away from the BPS model by including a small contribution from the standard term quadratic in derivatives might resolve this issue and yield the required small binding energies, but initial numerical computations \cite{Gillard2015} suggest that this is a rather singular perturbation.

Moving to an extreme parameter regime in the standard Skyrme model, where the quartic Skyrme term dominates over the usual leading order quadratic term, allows a potential term to significantly influence the properties of Skyrmions \cite{Harland}. This has been exploited by making a non-standard choice of the potential term to produce a lightly bound Skyrme model \cite{Gillard2015} with binding energies of the correct order of magnitude to model nuclei. Multi-Skyrmions in this model are very different from those in the standard Skyrme model and consist of portions of a lattice of well-separated single Skyrmions. Low binding energies are obtained because it is energetically unfavourable for Skyrmions to merge and indeed they can be effectively replaced by point particles \cite{Gillard2017}, so some of the nice solitonic features of Skyrmion models of nuclei become redundant. Whether the substantial suppression of the standard kinetic term for pions, used in both the lightly bound and BPS models, has serious implications for the description of pions within these models has yet to be fully investigated.

The theory that is closest in spirit to that in the present paper is the holographic model of Sakai and Sugimoto \cite{SS} and indeed the derivation \cite{Sut} of the rho meson extension of the standard Skyrme model may be viewed as a truncation of a flat space analogue of a holographic construction.
Baryons in the low energy effective action of the Sakai-Sugimoto model correspond to solitons in a bulk five-dimensional curved spacetime and the complexity of this model means that only the single soliton has been computed so far \cite{BolSut}. However, approximations suggest \cite{Bol1} that there may be regimes in which this model has multi-Skyrmions composed of well-separated single Skyrmions with similarities to those in the lightly bound model. It has also been shown \cite{Bol2} that there are parameter regimes where integrating out fields of the Sakai-Sugimoto model produces a generalized Skyrme model in which the dominant term is the one of sixth order in derivatives. These results hint at the possibility that all the apparently disparate recent attempts to modify the Skyrme model to reduce binding energies may in fact be related to each other in highly non-trivial ways.

\section{Including rho mesons in the Skyrme model}\quad
The standard Skyrme model is a nonlinear theory of pions in which the pion fields $(\pi_1,\pi_2,\pi_3)$ are combined into
the Skyrme field $U\in SU(2)$ as
\be
U=\begin{pmatrix}
\sigma +i\pi_3 & i\pi_1+\pi_2 \\ i\pi_1-\pi_2 & \sigma -i\pi_3
\end{pmatrix},
\ee
and the sigma field imposes the constraint $\sigma^2+\pi_1^2+\pi_2^2+\pi_3^2=1.$
The static energy is written in terms of the three
 $\mathfrak{su}(2)$-valued currents  
 $R_i=\partial_iU\,U^{-1}$ as
\be
E_\pi=\int \bigg(
-\frac{c_1}{2}\mbox{Tr}(R_iR_i)-\frac{c_2}{16}\mbox{Tr}([R_i,R_j]^2)
\bigg)\,d^3x,
\label{enpion}
\ee
where $c_1$ and $c_2$ are positive constants with values related to the choice of energy and length units. Baryons are described by Skyrmions, which are topological soliton solutions of the theory, with baryon number identified with the integer-valued topological charge
\be
B=-\frac{1}{24\pi^2}\int\varepsilon_{ijk}\mbox{Tr}(R_iR_jR_k)\, d^3x.
\ee
The Faddeev-Bogomolny energy bound \cite{Fa} for the Skyrme model is
\be
E_\pi\ge 12\pi^2\sqrt{c_1c_2}\,|B|,
\label{fadbog}
\ee
and the energy of the single Skyrmion exceeds this bound by over $20\%$, thereby providing plenty of room for large binding energies for Skyrmions with $B>1.$ 

The standard Skyrme model can be extended to a theory of pions and rho mesons by applying a dimensional deconstruction of pure Yang-Mills theory with an extra spatial dimension \cite{Sut}. The starting point is the Yang-Mills energy in
$\mathbb{R}^4$ given by
\be
E_{\mbox{\tiny{YM}}}=-\frac{1}{8}\int \mbox{Tr}(F_{IJ}F_{IJ})\,d^4x,
\label{yme}
\ee
where $x_I,$ with $I=1,..,4,$ denote the spatial coordinates
and  
$F_{IJ}=\partial_I A_J-\partial_J A_I+[A_I,A_J]$ are the components
of the $\mathfrak{su}(2)$-valued field strength.
The instanton number $N\in\mathbb{Z}$ of the gauge field provides the lower bound on the energy
\be
E_{\mbox{\tiny{YM}}}\ge 2\pi^2\, |N|,
\label{ymbound}
\ee
that is attained by fields that are either self-dual or anti-self dual. 

To deconstruct the fourth dimension we generalize the approach in \cite{Sut}
by writing $x_4=z$, fixing the gauge $A_z=0$ and writing the remaining three components in terms of the Skyrme currents $R_i$ and the $\mathfrak{su}(2)$-valued rho meson fields $\rho_i.$ Explicitly, we assume the restricted form
\be
A_i=-\frac{1}{2}(1+\phi)R_i+\alpha\phi'\rho_i,
\label{ansatz}
\ee
where $\phi(z)$ is a real-valued odd function satisfying the boundary condition $\phi(\infty)=1$ and $\alpha$ is a positive normalization constant.
Substituting this restricted form into the Yang-Mills energy (\ref{yme}), and integrating out the fourth dimension, the energy $E_{\mbox{\tiny{YM}}}$ defines a theory of pions and rho mesons with the energy given by
$E_{\pi,\rho}=E_\pi+E_\rho+E_{\rm int}$, where
\be
E_\rho=\int -\mbox{Tr}\bigg\{
\frac{1}{8}c_8(\partial_i \rho_j-\partial_j \rho_i)^2
+\frac{1}{4}m^2\rho_i^2
+c_3(\partial_i \rho_j-\partial_j \rho_i)[\rho_i,\rho_j]
+c_4[\rho_i,\rho_j]^2
\bigg\}\,d^3x,
\label{enrho}
\ee
and the interaction energy between the pions and rho mesons is
\bea
& &E_{\rm int}=\int -\mbox{Tr}\bigg\{
c_5([R_i,\rho_j]-[R_j,\rho_i])^2
-c_6[R_i,R_j](\partial_i \rho_j-\partial_j \rho_i)
-c_7[R_i,R_j][\rho_i,\rho_j]\nonumber\\
& &
+\frac{1}{2}c_6[R_i,R_j]([R_i,\rho_j]-[R_j,\rho_i])
-\frac{1}{8}c_8([R_i,\rho_j]-[R_j,\rho_i])(\partial_i \rho_j-\partial_j \rho_i)
\nonumber\\
& &-\frac{1}{2}c_{3}([R_i,\rho_j]-[R_j,\rho_i])[\rho_i,\rho_j]
\bigg\}\,d^3x.
\label{enint}
\eea
The constant coefficients in the above energy formulae are given by the integral expressions
\bea
c_1&=&\frac{1}{8}\int_{-\infty}^\infty \phi'^2\,dz, \quad
c_2=\frac{1}{8}\int_{-\infty}^\infty (1-\phi^2)^2\,dz, \quad
c_3=\frac{\alpha^3}{4}\int_{-\infty}^\infty \phi'^3\,dz, \quad
c_4=\frac{\alpha^4}{8}\int_{-\infty}^\infty \phi'^4\,dz, \quad \ \nonumber \\
c_5&=&\frac{\alpha^2}{32}\int_{-\infty}^\infty (1+\phi^2)\phi'^2\,dz, \quad
c_6=\frac{\alpha}{16}\int_{-\infty}^\infty (1-\phi^2)\phi'\,dz=\frac{\alpha}{12},  \nonumber\\
c_7&=&\frac{\alpha^2}{16}\int_{-\infty}^\infty (1-\phi^2)\phi'^2\,dz,  \quad
c_8=\alpha^2\int_{-\infty}^\infty \phi'^2\,dz, \quad
m^2=\alpha^2\int_{-\infty}^\infty \phi''^2\,dz.
\label{consts}
\eea
We fix the value of $\alpha$ by requiring the standard normalization $c_8=1$ for the term in the rho meson energy that is quadratic in derivatives.

The instanton number of the gauge field (\ref{ansatz}) is equal to the baryon number of the Skyrme field, that is $N=B$, hence the theory with pions and rho mesons inherits the Yang-Mills energy bound (\ref{ymbound}) to give
\be
E_{\pi,\rho}\ge 2\pi^2|B|.
\label{ymbound2}
\ee

Motivated by the requirement of obtaining a natural extension to include an arbitrary number of vector mesons, the choice of function made in previous studies \cite{Sut,Sut2} is 
\be
\phi(z)=\mbox{erf}(z/\sqrt{2}), 
\ee
and then $c_8=1$ requires that $\alpha={\pi^{1/4}}/{\sqrt{2}}.$
This yields values for the constants that we denote by Set I and can be found in Table~\ref{table:consts}, where some of these constants can be calculated analytically whereas others can only be found by computing the integrals in (\ref{consts}) numerically.
\begin{table}
\begin{center}  
  \begin{tabular}{|c | c c | c c |}
    \hline
  & \multicolumn{2}{c|}{Set I} & \multicolumn{2}{c|}{Set II} \\
 & exact & numerical & exact & numerical\\ 
  \hline
  $c_1$ & $\frac{1}{4\sqrt{\pi}}$ & 0.141 & $\frac{1}{6}$ &0.167\\[0.1cm]
    \hline
    $c_2$ & - & 0.198 & $\frac{1}{6}$ &0.167\\[0.1cm]
      \hline
      $c_3$ & $\frac{1}{2\sqrt{6}\pi^\frac{1}{4}}$ & 0.153 & $\frac{\sqrt{3}}{10}$ &0.173\\[0.1cm]
        \hline
        $c_4$ & $\frac{1}{8}\sqrt{\frac{1}{2\pi}}$ & 0.050 & $\frac{9}{140}$ &0.064\\[0.1cm]
          \hline
  $c_5$ & - & 0.038 & $\frac{3}{80}$ &0.038\\[0.1cm]
  \hline
  $c_6$ & $\frac{\pi^{1/4}}{12\sqrt{2}}$ & 0.078 & $\frac{\sqrt{3}}{24}$ &0.072\\[0.1cm]
  \hline
  $c_7$ & - & 0.049 & $\frac{1}{20}$ &0.050\\[0.1cm]
  \hline
  $c_8$ & 1 & 1.000 & 1 &1.000\\[0.1cm]
    \hline
    $m$ & $\frac{1}{\sqrt{2}}$ & 0.707 & $\frac{2}{\sqrt{5}}$ &0.894 \\[0.1cm]
    \hline
\end{tabular}
\end{center}
\caption{The values of the constants for Set I and Set II, where exact values are given if the required integrals can be calculated analytically and numerical values correspond to either the numerical evaluation of the exact result or the numerical computation of the integral if an exact result is unavailable.
}
    \label{table:consts}  
\end{table}
For Set I the Yang-Mills derived energy bound (\ref{ymbound2}) for the theory with pions and rho mesons is not as strict as the Faddeev-Bogomolny bound (\ref{fadbog}) for the pion theory alone as $2\pi^2 < 12\pi^2\sqrt{c_1c_2}= 2.005\times \pi^2.$

A motivation for a new parameter set is to obtain an extension of the standard Skyrme model in which the Faddeev-Bogomolny bound survives the inclusion of rho mesons by virtue of the fact that it coincides exactly with the Yang-Mills derived energy bound. The following simple application of the Cauchy-Schwarz inequality proves that the Yang-Mills derived bound cannot improve upon the Faddeev-Bogomolny bound,
\bea
12\pi^2\sqrt{c_1c_2}
&=&\frac{3\pi^2}{2}\sqrt{\bigg(\int_{-\infty}^\infty \phi'^2\,dz\bigg)
  \bigg(\int_{-\infty}^\infty (1-\phi^2)^2\,dz\bigg)}
 \nonumber\\
&\ge& \frac{3\pi^2}{2}\int_{-\infty}^\infty \phi'(1-\phi^2)\,dz
=
\frac{3\pi^2}{2}\int_{-1}^1 (1-\phi^2)\,d\phi=2\pi^2.
\eea
Saturation of the Cauchy-Schwarz inequality used in the above argument requires that there exists a positive constant $\beta$ such that
$
\phi'=\beta(1-\phi^2).
$
The Yang-Mills derived bound therefore agrees with the Faddeev-Bogomolny bound only if this equation is satisfied, which requires that 
$
\phi=\tanh(\beta z).
$
The scale $\beta$ controls the ratio of the coefficients in the Skyrme model, $c_1/c_2=\beta^2$, which we set to unity to match with traditional Skyrme units.
The new parameter set is therefore obtained from the function
\be \phi(z)=\mbox{tanh}(z),\ee
and the normalization constant is $\alpha={\sqrt{3}}/{2}$ to obtain $c_8=1.$
This yields the constants denoted by Set II in Table~\ref{table:consts},
where this time all constants can be calculated analytically. Note that despite the change of function there is still a reasonable agreement between the values of the constants in Set I and Set II.

In Section \ref{sec:results} we present the results of numerical computations to calculate Skyrmions and their binding energies using both parameter Set I and Set II. Before this, we show that within our framework it is impossible to obtain a parameter set in which rho mesons can be represented by a massive Yang-Mills field.

Traditionally, rho mesons are treated as a massive Yang-Mills field using an energy of the form
\be
E_\rho=\int-\mbox{Tr}\bigg\{\frac{c_8}{8}(\partial_i\rho_j-\partial_j\rho_i+g[\rho_i,\rho_j])^2+\frac{m^2}{4}\rho_i^2\bigg\}\,d^3x,
\label{gaugeform}
\ee
where $g$ is the gauge coupling constant. A comparison of this expression and
(\ref{enrho}) shows that a necessary condition for the Yang-Mills generated rho meson energy to have this form is that  $c_4c_8=2c_3^2$. However,
\be
\frac{8}{\alpha^6}(c_4c_8-2c_3^2)=
\bigg(\int_{-\infty}^\infty \phi'^4\,dz\bigg)
\bigg(\int_{-\infty}^\infty \phi'^2\,dz\bigg)
-\bigg(\int_{-\infty}^\infty \phi'^3\,dz\bigg)^2
>0
\ee
by the Cauchy-Schwarz inequality and the fact that $\phi(z)$ cannot be a linear function of $z$ as this is incompatible with the boundary conditions $\phi(\pm\infty)=\pm 1.$ This simple argument proves that within our framework there is no parameter set for a massive Yang-Mills formulation as ${2c_3^2}/{(c_4c_8)}<1$, but both parameter Set I and II are reasonably close to this unattainable limit, with
${2c_3^2}/{(c_4c_8)}={2\sqrt{2}}/{3}=0.94$ for Set I and
${2c_3^2}/{(c_4c_8)}={14}/{15}=0.93$ for Set II.

\section{Numerical results for Skyrmions}\label{sec:results}\quad
To obtain Skyrmions that minimize the energy $E_{\pi,\rho}$ we perform parallel numerical computations using a simulated annealing code with $130^3$ lattice points and spatial derivatives evaluated using second order finite difference approximations. The lattice spacing is $dx=0.08$ for parameter Set I and $dx=0.068$ for parameter Set II, so that the lattice spacing in Skyrme units is the same for both parameter sets.
At the boundary of the grid we fix the Skyrme field to be the identity matrix and all the components of the rho meson fields are set to zero. 
To reduce computational time we employ a course lattice for the initial phase of the minimization, with $65^3$ lattice points and a lattice spacing double the value on the fine lattice. We interpolate the fields from the coarse lattice to provide an initial field for energy minimization on the fine lattice. Applying a standard finite difference approximation on such a coarse lattice is inappropriate as the resolution is not fine enough to preserve the baryon number, but we avoid this problem by using Ward's logarithmic approximation \cite{Ward1995} for the sigma model energy of the Skyrme field. In fact, we find that not only does this method preserve the topology on our coarse lattice, but it also provides a surprisingly good estimate for the final energy obtained following interpolation and minimization on the fine lattice.  
\begin{figure}[ht]\begin{center}           
    \includegraphics[width=0.8\columnwidth]{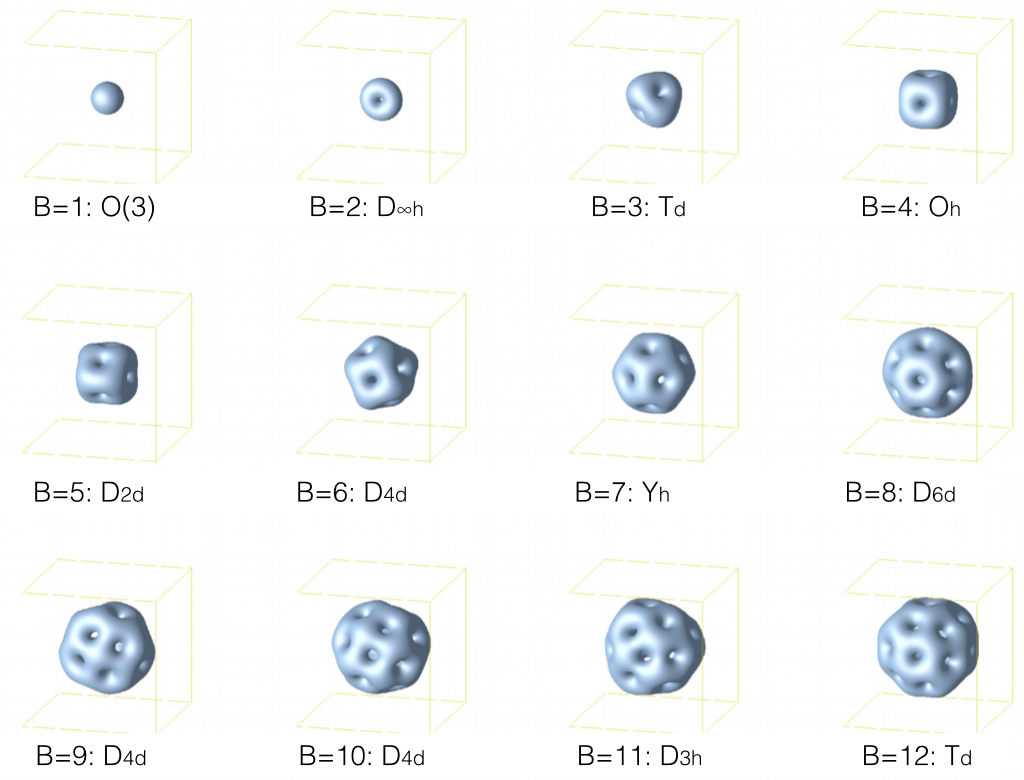}
    \caption{Energy density isosurfaces for Skyrmions with baryon numbers up to 12, in the model with pions and rho mesons using parameter Set I. The symmetry group of each Skyrmion is also listed.}
\label{fig:iso}
\end{center}\end{figure}
\begin{table}
 \begin{center} 
   \begin{tabular}{|c| c| c c|}
     \hline
 &  &  Set I & Set II\\
  $B$ & \large{$\frac{E_\pi}{2\pi^2 B}$} &  \multicolumn{2}{c|}{\large{$\frac{E_{\pi,\rho}}{2\pi^2 B}$}}\\[0.1cm]
\hline
1 & 1.2461 &  1.0624 &  1.0649\\
2 & 1.1912 &  1.0475 & 1.0485 \\
3 & 1.1566 &  1.0372 & 1.0375 \\
4 & 1.1300 &  1.0286 & 1.0284 \\
5 & 1.1274 &  1.0288 & 1.0284\\
6 & 1.1180 &  1.0260 & 1.0253\\
7 & 1.1043 &  1.0216 & 1.0206\\
8 & 1.1065 &  1.0233 & 1.0222\\
9 & 1.1057 &  1.0234 & 1.0221\\
10 & 1.1029 & 1.0225 & 1.0212\\
11 & 1.1027  & 1.0231 & 1.0217\\
12 & 1.0994  & 1.0225 & 1.0209\\
\hline
\end{tabular}
\caption{
The energy per baryon (in units of $2\pi^2$) for Skyrmions with
$1\le B\le 12$ in the Skyrme model ($E_\pi/(2\pi^2 B)$) and the model including rho mesons ($E_{\pi,\rho}/(2\pi^2 B)$) using parameter Set I and Set II.}
\label{table:energies}
\end{center}
\end{table}
\begin{figure}[ht]\begin{center}           
    \includegraphics[width=0.5\columnwidth]{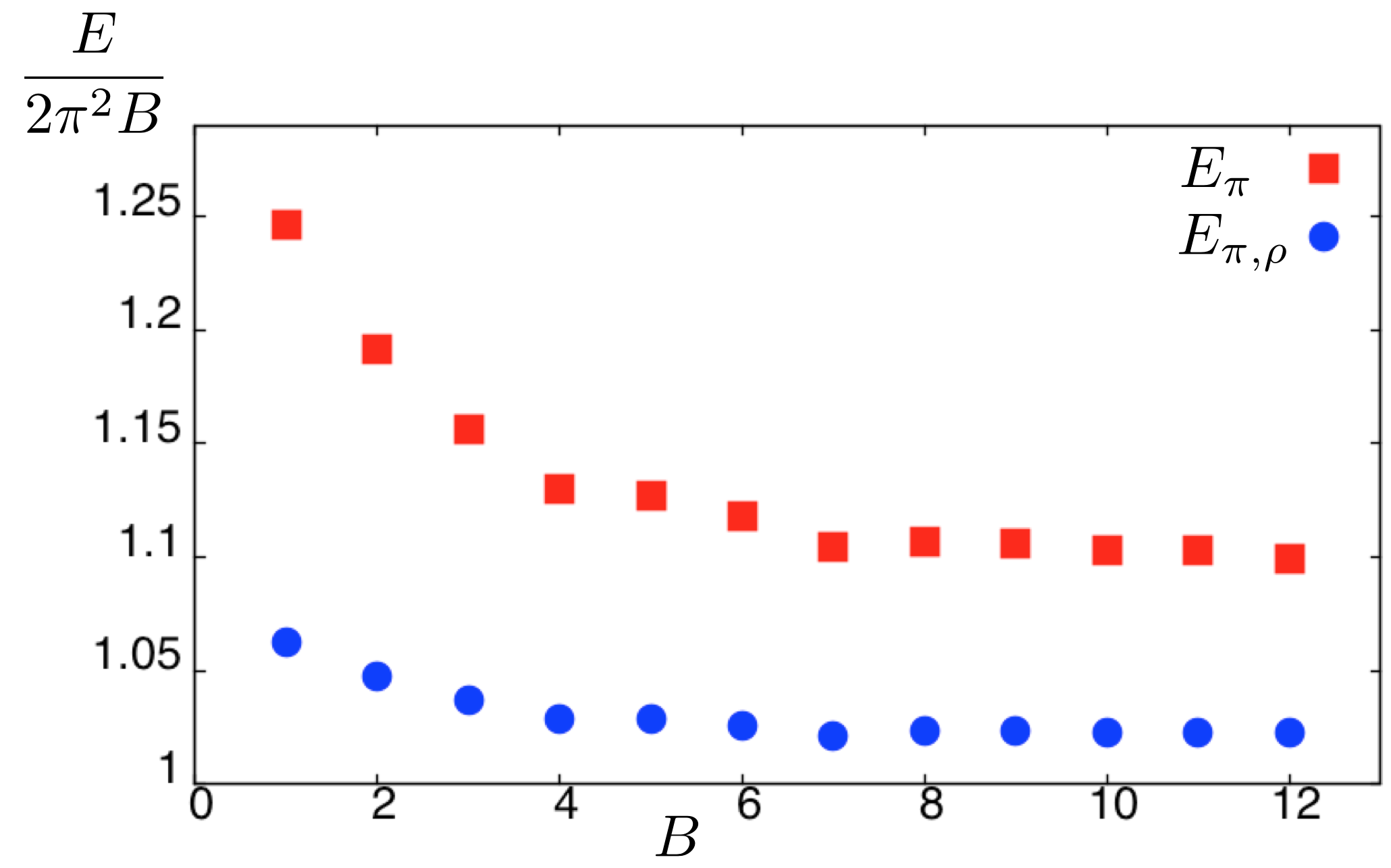}
    \caption{The energy per baryon (in units of $2\pi^2$) for Skyrmions with
      baryon number 1 to 12 in the standard Skyrme model (red squares) and the model including rho mesons (blue circles) using parameter Set I.
      }
\label{fig:en}
\end{center}\end{figure}

\begin{figure}[!hb]\begin{center}           
    \includegraphics[width=0.45\columnwidth]{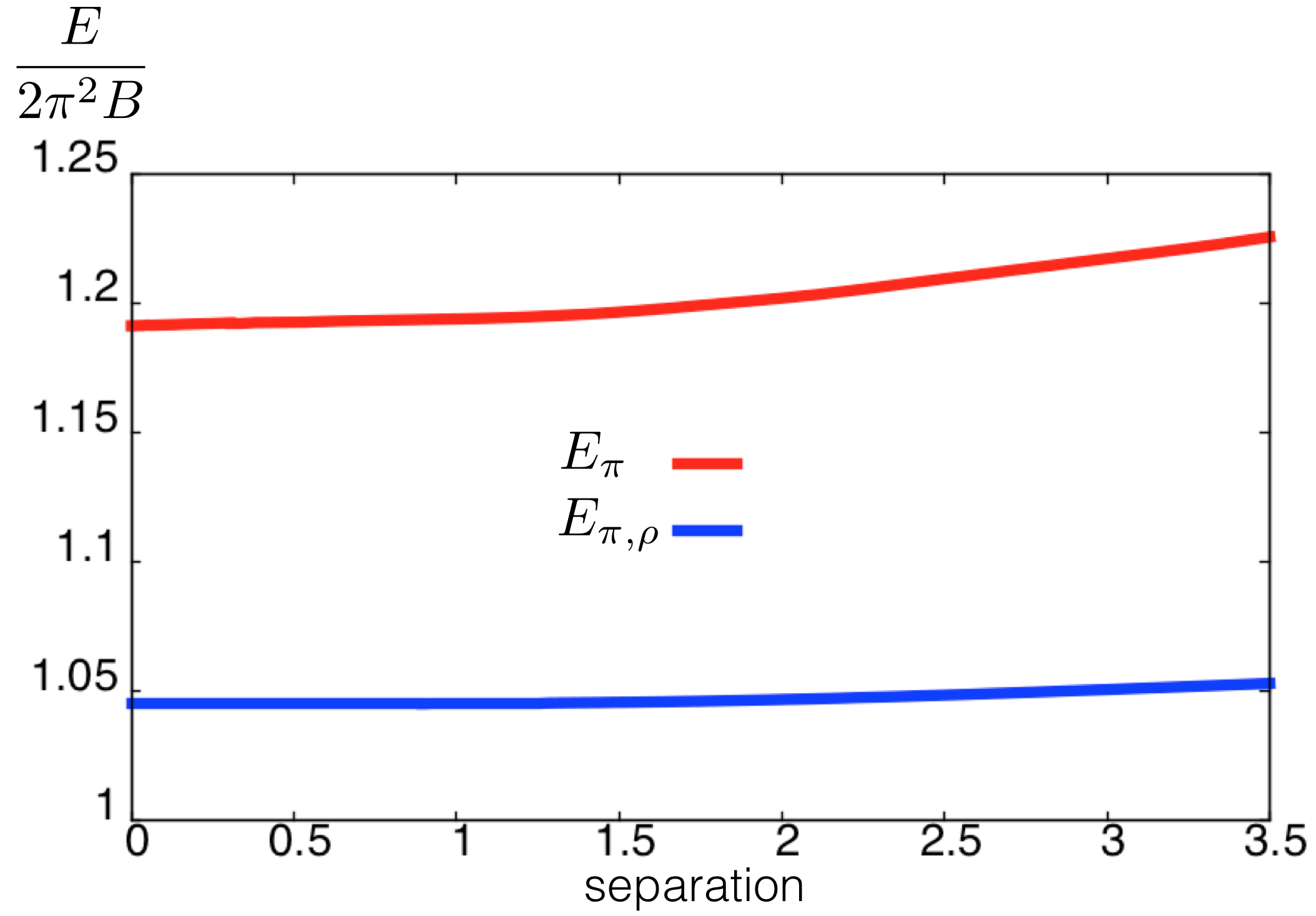}
    \caption{The energy per baryon (in units of $2\pi^2$) as a function of the separation between a pair of single Skyrmions in the standard Skyrme model (upper red curve) and the model including rho mesons (lower blue curve)  using parameter Set I.
      }
\label{fig:dist}
\end{center}\end{figure}
All initial conditions consist of vanishing rho meson fields, with the Skyrme field constructed from a range of methods that include the rational map ansatz \cite{HMS} and the product ansatz using both well-separated individual Skyrmions and clusters. For each baryon number, the Skyrmion we present is the lowest energy solution found using this variety of initial conditions, although other local energy minima were also obtained. We also apply the same procedure and algorithm to the standard Skyrme model to recompute the Skyrmions that minimize the energy $E_\pi$, to provide an accurate comparison between energies in the models with and without rho mesons.  

Figure~\ref{fig:iso} displays energy density isosurfaces for Skyrmions with
$1\le B\le 12$ using parameter Set I. The symmetry group of each Skyrmion is also shown and in all cases it is found to coincide with the symmetry group of the Skyrmion in the standard Skyrme model. This figure appears to show that the inclusion of rho mesons has had very little influence on Skyrmions. However, this overlooks the fact that there has been a significant reduction in the energies of these Skyrmions, as revealed in Table~\ref{table:energies}, where we present the energy per baryon for these Skyrmions and compare them with the values in the standard Skyrme model. We also show the same information in graphical form in Figure~\ref{fig:en}.

These results show that the inclusion of rho mesons has shifted Skyrmion energies much closer to the lower bound and has reduced the binding energies of these Skyrmions from around $15\%$ to less than $4\%.$ Although the inclusion of rho mesons has not decreased binding energies to the $1\%$ level seen in experimental data, this reduction is a big step towards more realistic values and demonstrates that the inclusion of the heavier mesons neglected in the standard Skyrme model is a feasible mechanism to move Skyrmion theory closer to reality. Furthermore, unlike other recent attempts to modify the Skyrme model to lower binding energies, this is achieved without any change in the qualitative features of Skyrmions.

In Table~\ref{table:energies} we also present Skyrmion energies for parameter Set II, where we see that the results are very similar. In fact, plots of energy density isosurfaces for Skyrmions with parameter Set II are indistinguishable from those presented for Set I in Figure~\ref{fig:iso}, so they are not shown here. The reduction of binding energies to around $4\%$ seems to be a rigid result in theories of this type.

As one might expect, a reduction in binding energies is accompanied by a flattening of the interaction potential between Skyrmions. This is demonstrated in Figure~\ref{fig:dist} where we plot the energy as a function of separation between two single Skyrmions for the standard Skyrme model (upper curve) and the model of pions and rho mesons using parameter Set I (lower curve). This data is calculated by tracking the positions of two initially separated single Skyrmions during the energy minimizing evolution. Recent work \cite{Halcrow2016,Halcrow2017} has highlighted the importance of vibrational modes in ordering the spin states of quantized Skyrmions, and the softening of these modes obtained by the addition of rho mesons will make these considerations even more important.

The flattening of the interaction potential mirrors the results found for the Skyrme crystal with pions and rho mesons \cite{Ma}, where an approximate description in terms of Fourier modes reveals only a weak dependence of the energy per baryon on the period of the crystal. Furthermore, the minimal energy per baryon corresponds to a binding energy that is consistent with the results presented in the present paper for finite baryon numbers. 

\section{Conclusion}\quad
We have considered Skyrmions in models of pions and rho mesons obtained by integrating out the extra dimension of pure Yang-Mills theory. These models have an advantage over traditional attempts to incorporate rho mesons, in that a topological energy bound survives the inclusion of vector mesons and it can even be arranged so that the same Faddeev-Bogomolny bound of the standard Skyrme model survives this extension. By applying parallel numerical computations we have obtained the first multi-Skyrmions in theories of this type and shown that binding energies are dramatically reduced. Skyrmion energies are much closer to the topological lower bound than in the standard Skyrme model, despite very little change in the qualitative features of Skyrmions. It is expected that including the next heaviest mesons will further reduce binding energies but it will be a significant numerical challenge to include these extra fields because of the increase in the number of degrees of freedom and the number of terms that need to be evaluated in computing the energy.

The studies in this paper have considered massless pions but including a pion mass term would also be interesting and is expected to change the qualitative features of Skyrmions for large enough baryon numbers, as in the case of the standard Skyrme model \cite{BatSut}. It might also be interesting to investigate the pion potential used in the lightly bound model \cite{Gillard2015}, to see whether it can have such significant consequences on the structure of Skyrmions when the influence of the term quadratic in pion field derivatives is suppressed not by reducing the coefficient of this term but by interactions between pions and rho mesons.

Using the expression for the energy $E_{\pi,\rho}$ given in this paper one could ignore its origin from Yang-Mills theory and simply consider this theory without the requirement that the parameters are constrained by the integral expressions (\ref{consts}). Of course, the topological lower bound is then lost and it is not even obvious what constraints need to be imposed on the parameter set so that the energy remains positive. One way to attempt to simplify the parameter space is to enforce a massive Yang-Mills field formulation for the rho mesons. We have made some preliminary investigations of this approach by perturbing away from the parameters given by Set I and Set II to enforce the required relations between the parameters required for a gauge field formulation but initial results suggest that the original parameter sets are close to optimal in terms of reducing binding energies. As the parameter space is large it is not possible to make a systematic study of Skyrmion energies in theories of this type, so it is possible that an improved parameter set exists, but without a sophisticated derivation of a parameter set it seems unlikely that one could be found simply by brute computation.

The results in this paper are presented in a form that is independent of the calibration of the model, namely the specification of physical energy and length units. In the standard Skyrme model there are many ways to calibrate the model, with the original calibration obtained by matching the masses of the nucleon and delta resonance using a rigid rotor quantization \cite{ANW,AN}. However, the delta is a broad resonance that strongly radiates pions, which is related to the fact that the spin of the delta strongly deforms the $B=1$ Skyrmion and this has a rather complicated effect on the calibration of the model \cite{BKS}. Furthermore, the calibration using the $B=1$ sector does not provide an accurate fit for the properties of nuclei with larger values of $B$, so alternative calibrations have been proposed to provide a better match for larger nuclei \cite{MMW}. These same calibration issues remain for the extended Skyrme model, at a similar level to the standard Skyrme model, hence the reason that we present our results in a form that is independent of the choice of calibration.

\section*{Acknowledgements}
\noindent
We thank Matt Elliot-Ripley, David Foster, Nick Manton and Tom Winyard for useful discussions. This work is funded by the European Union Horizon 2020 research and innovation programme under the Marie Sk\l odowska-Curie grant agreement No 702329. The parallel computations were performed on Hamilton, the Durham University HPC cluster. 


\end{document}